# AstroEBSD: exploring new space in pattern indexing with methods launched from an astronomical approach

Authors

**Thomas Benjamin Britton[a]*, Vivian Tong[a], Jim Hickey[a], Alex Foden[a] and Angus Wilkinson[b]**

[a]Materials, Imperial College London, London, SW7 2AZ, UK

[b]Materials, University of Oxford, Oxford, OX1 3UJ, UK

Correspondence email: b.britton@imperail.ac.uk

**Funding information** Engineering and Physical Sciences Research Council (grant No. EP/H018921/1 to Angus Wilkinson, Thomas Benjamin Britton; grant No. EP/K034332/1 to Vivian Tong, Thomas Benjamin Britton, Angus Wilkinson); Royal Academy of Engineering (grant No. Research Fellowship to Thomas Benjamin Britton); Shell (contract No. Shell-Imperial Advanced Interfaces UTC).

**Synopsis** We introduce a method to automatically index electron backscatter diffraction patterns and demonstrate this with dynamical diffraction simulations and experimental patterns.

**Abstract** Electron backscatter diffraction (EBSD) is a technique used to measure crystallographic features in the scanning electron microscope. The technique is highly automated and readily accessible in many laboratories. EBSD pattern indexing is conventionally performed with raw electron backscatter patterns (EBSPs). These patterns are software processed to locate the band centres (and sometimes edges) from which the crystallographic index of each band is determined. Once a consistent index for many bands are obtained, the crystal orientation with respect to a reference sample & detector orientation can be determined and presented. Unfortunately, due to challenges related to crystal symmetry, there are limited available pattern indexing approaches and this has likely hampered open development of the technique. In this manuscript, we present a new method of pattern indexing, based upon a method with which satellites locate themselves in the night sky, and systematically demonstrate its effectiveness using dynamical simulations and real experimental patterns. The benefit of releasing this new algorithm as open source software is demonstrated as we utilise this indexing process, together with dynamical solutions, to provide some of the first accuracy assessments of an indexing solution. In disclosing a new indexing algorithm, and software processing tool-kit, we hope this opens up EBSD developments to more users. The software code and example data is released alongside this article for 3rd party developments.

**Keywords: Electron Kikuchi diffraction; Indexing; Astronomy.**

## 1. Introduction

The first electron backscatter diffraction patterns were captured by Naishikawa and Kikuchi (1928) using a cathode ray and a sample of calcite. These patterns were later explored for more samples by Alam *et al.* (1954)., where sharp diffraction patterns were captured on film mounted as a cylinder around the sample within a vacuum chamber. Subsequent developments in pattern capture and image analysis have moved the technique towards its place as a modern, and automated, staple within the characterisation toolbox. In general, the indexing of diffraction patterns has largely been handled by commercial software, with the limited release of information concerning the algorithm. However, notable exceptions have subsequently resulted in commercial software, such as the PhD thesis Wright (1992) and related publications from the Yale / BYU / TSL (now EDAX) research groups (Adams *et al.*, 1994, Kunze *et al.*, 1993, Wright & Adams, 1992). Together with work by Krieger-Lassen and colleagues (Lassen, 1996b, a, 1998, 1999, Lassen & Bildesorensen, 1993, Lassen *et al.*, 1992), this is likely the basis of the HKL (now Oxford Instruments toolkit), and the work of Schwarzer (Schwarzer, 1991a, b, 1993). However, specific detail and evolution of algorithms within commercial software are generally not divulged.

The EBSD method is now highly automated and widely established, typically through commercial systems, in laboratories world-wide. A recent review by Wilkinson and Britton (2012) highlights how rapid automation has enabled this technology to bolster our materials characterisation toolbox. In particular, we note that many users may take advantage of the automated indexing scheme to generate rich surface maps. However, complications often arise due to the presence of multiple reference frames, and through use of systems from many different manufactures. We have noticed that there are systematic differences in the origin of indexing noise, and the source of these errors are difficult to trace due to the closed nature of many of their indexing schemes. In part, this has motivated the present work.

In this paper, we will introduce key reference frames and equations required to transform electron backscatter patterns from real space, to Radon space, and to locate the band centres. Then we introduce a new indexing routine based upon an astronomical approach (originally suggested to the Authors by Dr Austin Day), which was originally developed by Groth (1986) for the Hubble Space Telescope and also used in whale shark identification (Arzoumanian *et al.*, 2005)). We adapt the algorithm for the case of crystal symmetry, which causes significant complications. Using this indexing method, we present a method to locate the pattern centre (assuming that strain is negligible) through an optimisation scheme to find the best 'fit' between the indexed bands and the look-up-table (similar to the work of (Zambaldi *et al.*, 2009) but refined for the absence of bands and optimisation of a wider search space). The potential to establish the fundamental accuracy of an indexing solution is realised through use of gnomonic projections of dynamical diffraction simulations (generated with the Bruker Dynamics software) of α-titanium (hexagonal closed packed, HCP). Ultimately, the utility of this indexing approach is demonstrated using experimental patterns from a silicon wafer (Diamond cubic, and close to the cubic closed packed, aka FCC structure) and a deformed α-iron (body centred cubic, BCC) polycrystal.

## 2. Software Processing

### 2.1. Background correction & flat fielding

A typical EBSD detector consists of a flat phosphor screen inserted within a microscope chamber (Figure 1), diffracted electrons form a pattern on the screen, which is imaged using a series of optical lenses and a CCD (or CMOS) imaging device. In a typical EBSD experiment, many electron backscatter patterns (EBSPs) are sequentially captured as the beam is rastered across the surface of the sample. To optimise contrast for indexing of these EBSPs, software based image processing is typically performed.

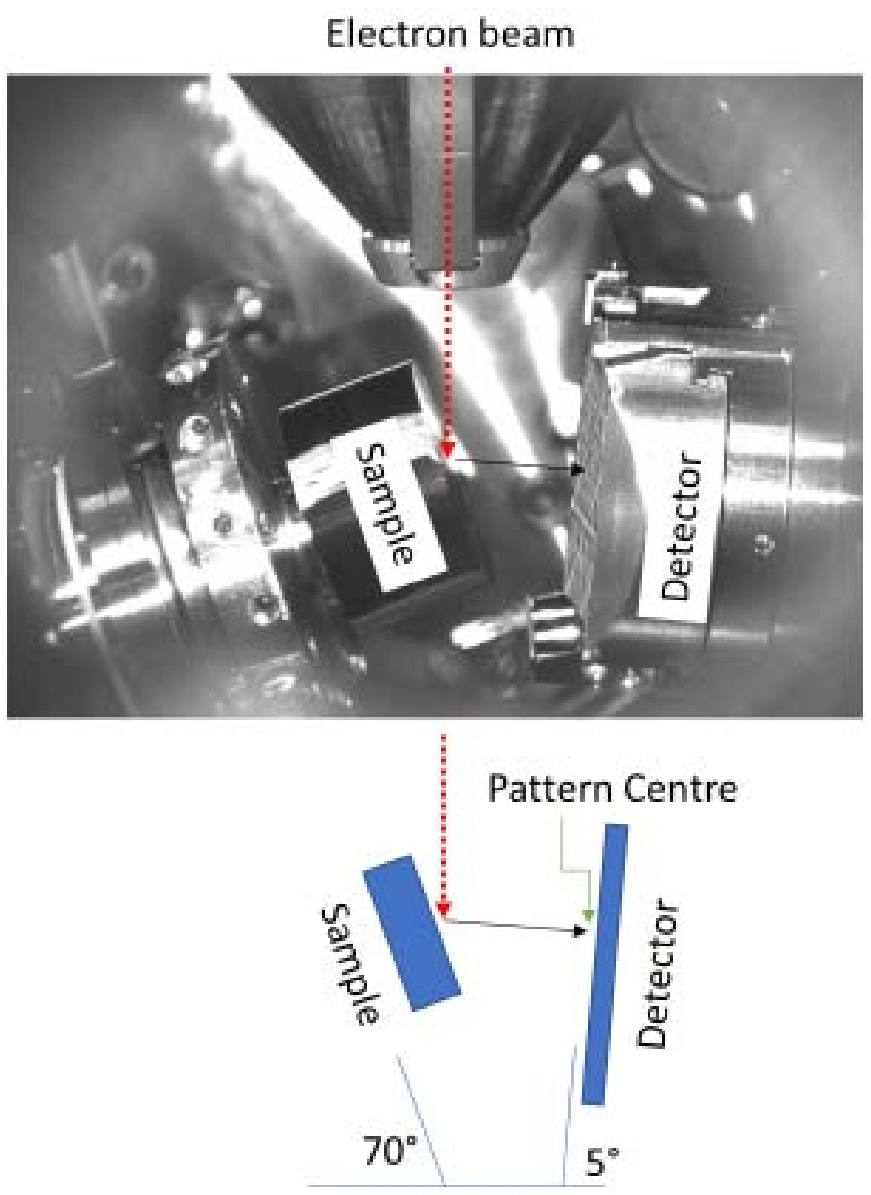

**Figure 1** Infrared photograph, with an EBSP superimposed, of an experimental set-up for conventional electron backscatter diffraction and associated cartoon demonstrating how the sample is highly tilted (typically to 70°) and the camera is inserted at a tilted angle (here ~5°).

The raw EBSP (Figure 2A) is stored in memory and consists of a 2D array of intensities, based upon the interaction of electrons with the EBSD detector phosphor screen. A raw EBSP contains two major features, namely a smoothly varying background and more localised Kikuchi bands. The background has its root in the superposition of the channelling in, compositional and topographical contrast that gives rise to a significant approximately Gaussian background signal. If the screen is largely undamaged the Kikuchi bands of raised intensities, arising from channelling out and the diffraction of electrons, are (broadly) only a fraction of this signal (~10%) and are localised due to the orientation of the crystal and the structure of the crystal lattice (i.e. atom positions, their relative contribution to the dynamical signal, the volume of the unit cell and the lattice parameters, and the quality of the crystal under the electron beam). For more details of how the bands within EBSPs are formed, the reader is directed towards the work of Winkelmann (2009).

To correctly index the diffraction pattern, it must be processed to optimise the contrast of the (useful) signal with respect to the background, and flat fielded to optimise the efficacy of the peak identification and pattern indexing scheme. In this work, we perform an initial pattern resize (keeping the aspect ratio of the pattern intact). Subsequently, we divide the image by a Gaussian blur of the initial image. The blur kernel is selected to be significantly wider than the bands in the pattern and this is a quick method of performing a high pass spatial frequency filter. The resultant EBSP is subsequently cropped to remove data towards the corner of the detector, using a circular mask, as vignetting in the optical capture system makes flat fielding difficult. Background correction can also be adapted to include other steps, if required, including feature removal (for instance if the screen is damaged) but this screen was in good condition, so further corrections are not explored here.

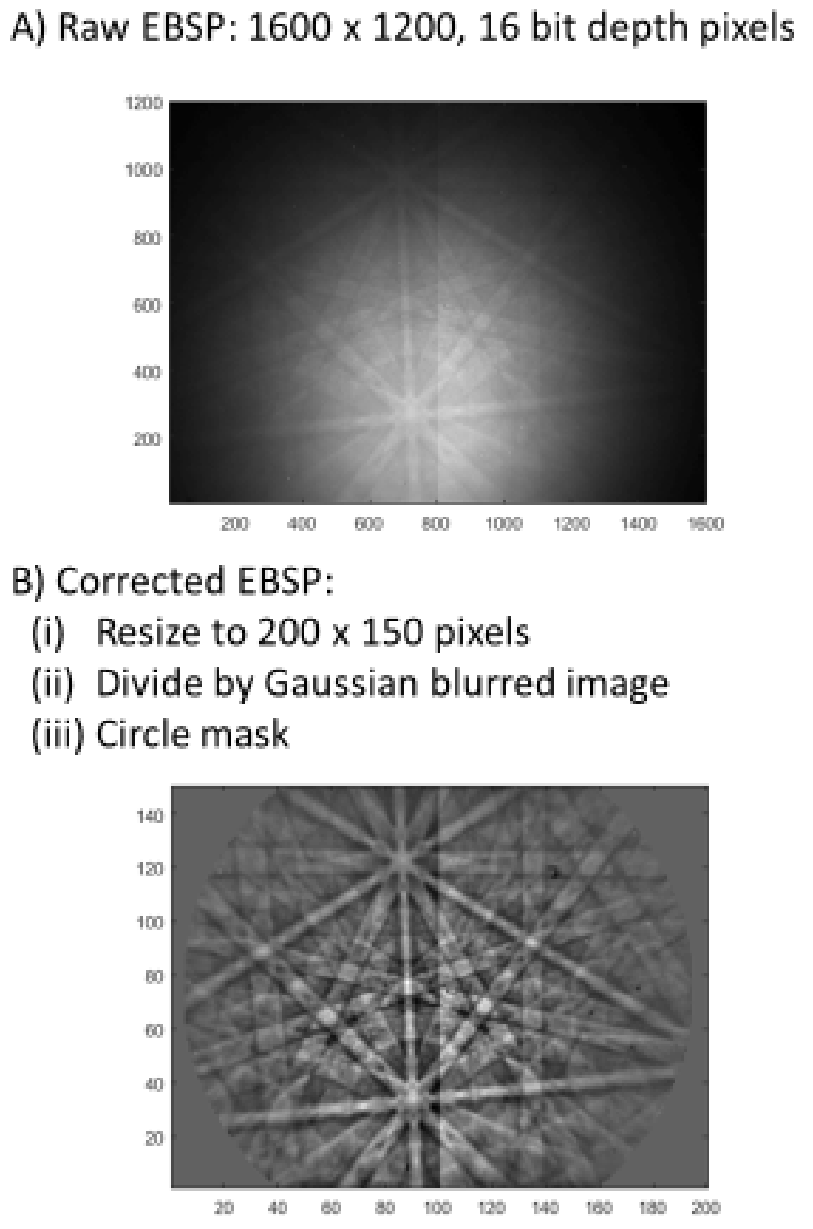

**Figure 2** : Software based background correction of raw electron backscatter pattern (EBSP) for indexing.

### 2.2. Radon transformation and peak localisation

We wish to locate the bands within the EBSP, as noted by the Riso group (Lassen *et al.*, 1992), we can utilise a Radon transformation to convert bands of raised intensity into points of raised intensity which are relatively easy to find using a computer. Formally each band edge is a hyperbolic Kossel cone and the centre line is perpendicular to the plane normal of the diffracting lattice plane (i.e. the majority of diffraction information within an EBSP is a direct projection of the lattice plane). In conventional geometries and voltages, the divergence of the two band edges is only a few degrees can be neglected, so we can utilise a parallel projection and integration of the pattern perpendicular to a sampling line. This is the Radon transformation. The Radon transformation is related directly to the Hough transform (which was formally only introduced as a binary transformation and later adjusted to a greyscale method (Hough, 1962, Duda & Hart, 1972)) which is more typically used in many, but not all, of the commercial EBSP analysis packages, largely for historical reasons.

Within Matlab, the Radon transform performs line based integrations about the image centre. The number of sampling lines, i.e. ρ-steps, is proportional to the greatest dimension of the image (e.g. a 400 x 300 image will result in 400 ρ-steps). The number of sampling angles is determined by the user, and this can often be 1°. The resolution of the Radon transform is linked to the resolution that the band can be localised to, and the computational cost of this step may be high for higher resolutions (or the transformation of larger images).

Furthermore, for EBSP analysis, it is useful to flatfield the Radon transform by a pattern mask. This resolves two major issues:

(1) the number of pixels in the path towards the edge of the screen is fewer than those near the centre, and so there is an expected variation in intensity;

(2) there are sampling artifacts when sampling horizontal, vertical, or at 45° angles to these (Tao & Eades, 2005).

For a sharp diffraction pattern of a crystal with centrosymmetry (and minimal anisotropic emission artifacts) the band edges are sharp and symmetrical (Winkelmann, 2009). This means a ‘dark-light-dark’ band profile transforms into a ‘dark-light-dark’ profile along the ρ direction within the Radon transform.

We choose to localise the peak centres with 2D convolution of the Radon transform (Figure 3B) with a Sobel based edge filter, operating along the $\rho$ direction:

$$\Psi = \begin{pmatrix} -1 & -2 & -1 \\ 0 & 0 & 0 \\ 1 & 2 & 1 \end{pmatrix} \qquad \text{Equation 1}$$

From this edge transform, a peak search is used to find the highest contrast bands within the edge map (Figure 3C), through searching of the positive and nearby negative gradient (confined within a window). The band centre is the mean of these two peaks. We introduce threshold values to optimise this peak searching to avoid very narrow ‘bands’ (e.g. the ‘split chip artefact’ in Figure 3A) and a maximum search width based upon the likely ‘2θ’ values based upon Bragg’s law for the experimental set-up. Alternative image processing strategies may involve use of a butterfly convolution mask or a top hat filter (Pinard *et al.*, 2009, Lassen *et al.*, 1992).

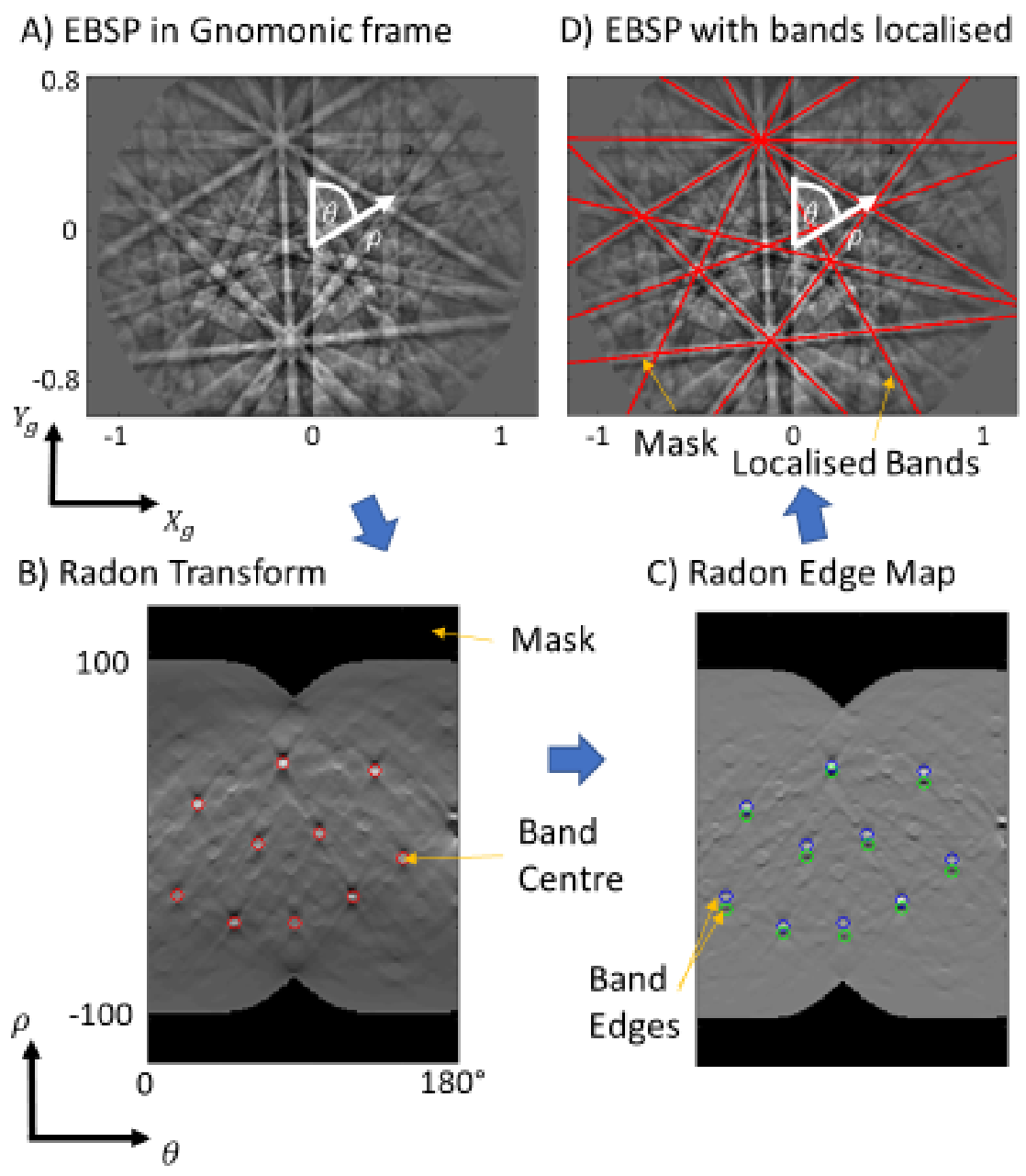

**Figure 3** Transformation of an EBSP for peak identification using edge detection in Radon space. (A) The resized pattern is presented in the Gnomonic frame, with the pattern centre at the (0,0). (B) Bands are identified using a Radon transform. The range of ρ values is given by the input pattern size. The range of θ values is selected by the user. Circular cropping (principally to manage image vignetting) results in a black mask. (C) The transformed image is edge filtered along ρ and the band edges are identified. (D) Bands detected in Radon space are overlaid on the input EBSP.

Now, we must introduce coordinate systems to link the band centres found within the Radon transform to those within the pattern. In Matlab, the Radon transform calculates parallel projection integrations for sampling lines of a fixed range of (clockwise) angles θ°. Each projection is taken along a line with a distance from the central point, ρ, as given in pixels. This is unlikely to be the pattern centre. The resolution of the Radon transform in ρ is determined with respect to the original image size (in this case it is ±100 x √2 pixels).

We can therefore relate the inscription of a band centre on the screen to the normal vector of the diffracting plane. For simplicity, the pattern is presented with respect to a gnomonic frame, such that the pattern centre is at [0,0] and the screen is at a distance Z=+1 (formal introduction of this gnomonic frame is given in a tutorial paper (Britton et al., 2016)), and in the screen coordinate system X points horizontally from 'west' to 'east' and Y points from 'bottom' to 'top' (see Figure 3A).

To calculate the plane normal vector, we introduce a transformation from image coordinates to the gnomonic space starting with three coordinate transformations $(T_x, T_y, T_s)$:

$$T_x = \frac{S_x}{S_y} \frac{\left(\frac{(S_x+1)}{2} - 1\right)}{S_x - 1} - \frac{PC_x}{PC_z} \quad \text{Equation 2}$$

$$T_y = \frac{\left(\frac{(S_y+1)}{2} - 1\right)}{S_y - 1} - \frac{1 - PC_y}{PC_z} \qquad \text{Equation 3}$$

$$T_s = \frac{1}{PC_z} * \frac{1}{S_x - 1} \qquad \text{Equation 4}$$

Where: $[S_x, S_y]$ is the screen size in pixels, $[PC_x, PC_y, PC_z]$ is the projection centre in fractions of the pattern width, height, and width respectively (as per conventions in (Britton *et al.*, 2016)),

If we have $[\rho, \theta]$ as the position of the band as identified within the Radon transform, the plane normal, $\boldsymbol{n}$, can be obtained from:

$$\boldsymbol{n} = \pm\left[0, -T_x + T_y \tan(\theta) - \frac{T_s\rho}{\cos(\theta)}, \tan(\theta)\right] \times [\tan(\theta), 1, 0] \qquad \text{Equation 5}$$

This vector is obtained from the cross product of the line of that crosses the screen (obtained from localisation within the Radon transform) and line that goes from the projection centre to a point on the line.

The length of this normal vector is unknown and not needed for indexing and so it is converted into a unit normal vector.

The plane normal can be back projected onto the screen using the Gnomonic reference frame (which is reported in more detail, with associated code, in Britton et al. (2016)) for visualisation and interpretation of each diffraction pattern.

## 3. Indexing

At this stage, we have located a number of bands of high intensity within our EBSP and related these to the normal vectors of the diffracting lattice planes. Next we determine the crystal family for each Kikuchi band, and subsequently find a consistent indexing that accommodates crystal symmetry.

It is well known that, regardless of the crystal orientation, the angles between crystal planes are invariant. Therefore, if we can relate the relative position of the plane normal, we can assign the appropriate crystallographic indices of each diffracting lattice plane.

The challenge of indexing each plane normal is equivalent to the challenge of identifying stars within the night sky, and this identification challenge is required in order to determine the relative orientation of a satellite that is moving through space. We can therefore translate solutions from the astronomy community toward solving the orientation of crystals, if we consider each plane normal as a 'star'. In the astronomical algorithms, presented by (Groth, 1986), a group of three stars can be used to generate 'a characteristic triangle' which encode the angular information about the relative positions of each star. Therefore, for a diffracting crystal, we can use knowledge of the crystal symmetry and active diffracting lattice planes, to calculate a pre-determined look up table (LUT).

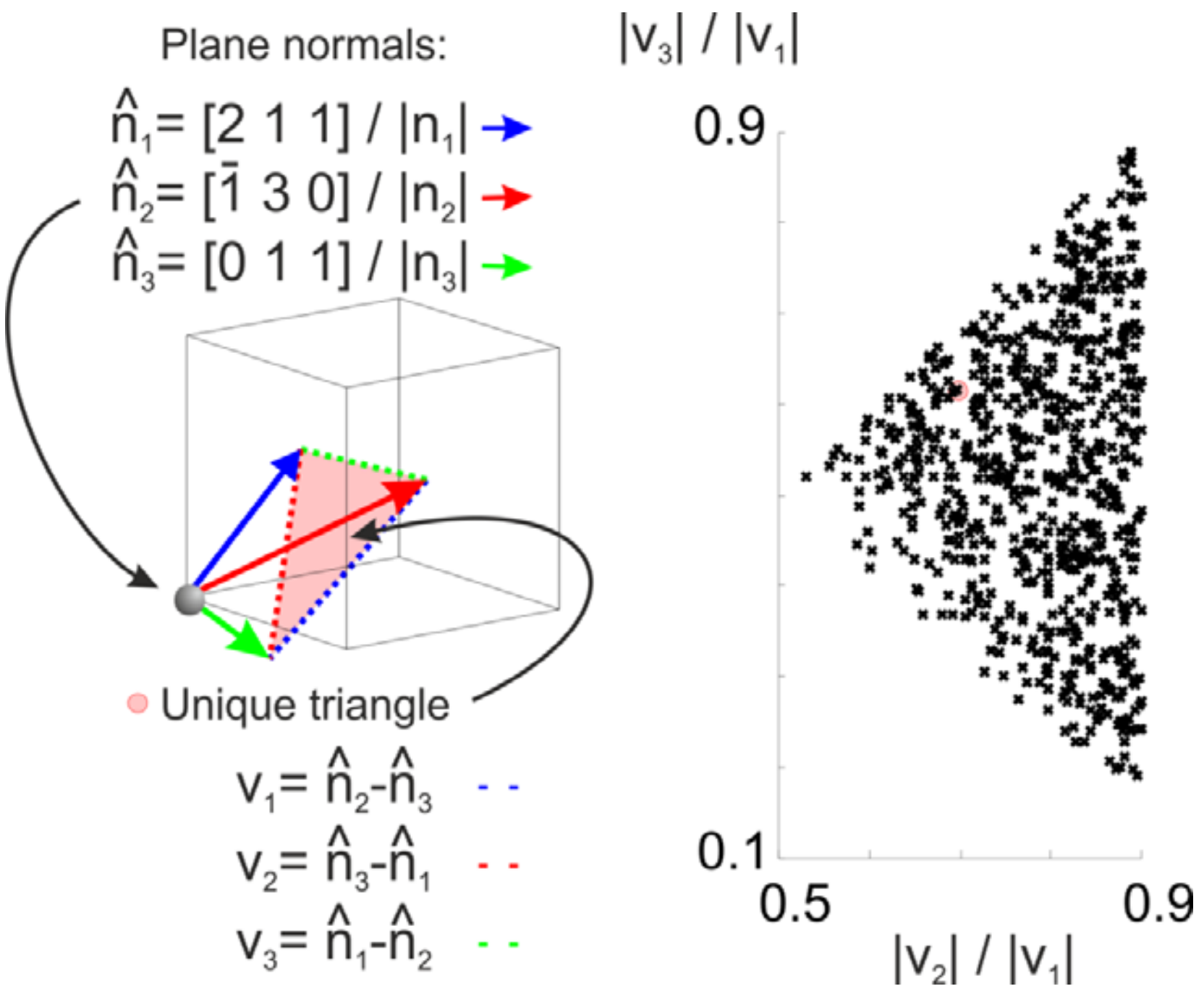

**Figure 4** Generation of a characteristic triangle for three example unit normal vectors from a diffraction plane normal (aka reflector) list and the corresponding look up table for the {110}, {200}, {211} and {310} reflectors. Note that the sides of the unique triangle are sorted such that $|\mathbf{V_1}| > |\mathbf{V_2}| > |\mathbf{V_3}|$. There is an outlined red circle within the look up table which corresponds to this unique triangle.

For each phase, generation of an appropriate LUT is performed, as shown in Figure 4 for a cubic crystal. In practice, the LUT contains the ratio of the edges of the triangle. We construct from a combination of three unit normal vector directions, $\hat{\boldsymbol{n}}$, and calculate three difference vectors, $\boldsymbol{v_i} = \widehat{\boldsymbol{n_j}} - \widehat{\boldsymbol{n_k}}$. The length the difference vectors are calculated and sorted by magnitude. The ratio of the shortest and second shortest lengths over the longest are calculated, i.e. $|\boldsymbol{v_3}|/|\boldsymbol{v_1}|$ and $|\boldsymbol{v_2}|/|\boldsymbol{v_1}|$ and these are displayed pictorially in Figure 4. In order to reduce ambiguity with triangles with sides of nearly equal length, those with $|\boldsymbol{v_2}|/|\boldsymbol{v_1}| > 0.9$ are removed from the look up table. To reduce computational cost and speed up accurate indexing, combinations of planes that have the same point within the LUT are encoded in a second 'sub-table' (this sub-table contains symmetric groups and band combinations that give rise to the same interplanar angle grouping). From a computational perspective, it is important that this look-up-table, and the sub-table, only need to be calculated once per phase.

For each pattern, we use an ‘n choose 3’ operation to generate groups of three identified bands and find potential matches within the LUT, using a ‘search length’ in the LUT, which describes the radius of the red circle in Figure 4**Error! Reference source not found.**. The radius of this circle could be adjusted to address indexing issues within a sparse or dense LUT. Upon finding a band, we note the potential plane family for each band in an accumulator array. All the characteristic triangles are compared in turn, and we identify the most trustworthy crystal family for each band (i.e. it has the most values against an individual plane family within the accumulator array).

After the most trustworthy crystal families are identified, we assign a consistent indexing solution for our crystal. This involves selecting one symmetry for an initial band and assigning a consistent frame of reference for all subsequently indexed diffracting bands.

In practice, we may have a band that is inconsistent with our indexing approach. This may happen for instance due to poor peak identification within our Radon accumulator, or the lack of a diffracting plane family within our LUT. Therefore, we try to find a best fitting solution that has the maximum number of indexed bands, where we use two bands to find the relative crystal orientation between test and reference, from which this rotation matrix can be computed to check if consistent indexing of the remainder of the bands within pattern can be found. Bands which do not match the orientation (i.e. are more than 3° from their expected plane normal direction) are discarded.

For the best indexing of the pattern, the orientation of the crystal is refined in calculating the crystal orientation matrix, **G**, from the Kabsch algorithm (Kabsch, 1976, 1978) to minimise the root mean square deviation between the plane normal vectors determined from the pattern and those within the LUT. In practice this means using a cross covariance matrix of indexed bands within the pattern and their respective bands within the LUT. This orientation matrix may be used to describe the crystal in the frame of the detector and can be used to index the pattern, as shown in Figure 5.

Once the best indexing is found, we calculate the Mean Angular Error (MAE) for all the successfully indexed bands:

$$|\theta| = |\arccos(n_{test}.\mathbf{G}n_{ref})| \qquad \text{Equation 6}$$

At this stage, the orientation is described using **G** which relates a reference crystal (used to generate the LUT) to the diffraction pattern. We can transform this orientation matrix into other frames of reference (e.g. the sample frame (Britton *et al.*, 2016)) or representations using other representations (e.g. quaternions & Euler angles) as required (Rowenhorst *et al.*, 2015).

**6 – Pattern Centre Determination**

Ideally, indexing of EBSPs requires knowledge of the source point position. However, it is possible to utilise an indexing algorithm to find the pattern centre with a reasonable precision for orientation analysis. An incorrect erroneously pattern centre will change the apparent interplanar angles, and therefore we can adjust the pattern centre until indexing is improved. A useful metric for assessing the quality of indexing is to calculate the mean angular error, which is the mean value of the magnitude of differences between the measured and expected interplanar angles averaged over all band pairs.

In practice when we need to measure the pattern centre, we calculate the mean angular error for each indexed solution and weight this with respect to a solution that utilises more bands. The best pattern centre is determined when this weighted mean angular error (WMAE) is minimised.

$$WMAE = (num\ bands\ sought - number\ bands\ found + 1) * MAE \qquad \text{Equation 7}$$

To validate this approach, we have used a dynamical pattern simulation with a known orientation and pattern centre and explored the pattern centre searching space. The successful indexing of an $\alpha$-titanium diffraction pattern, using the exact pattern centre is shown in Figure 5.

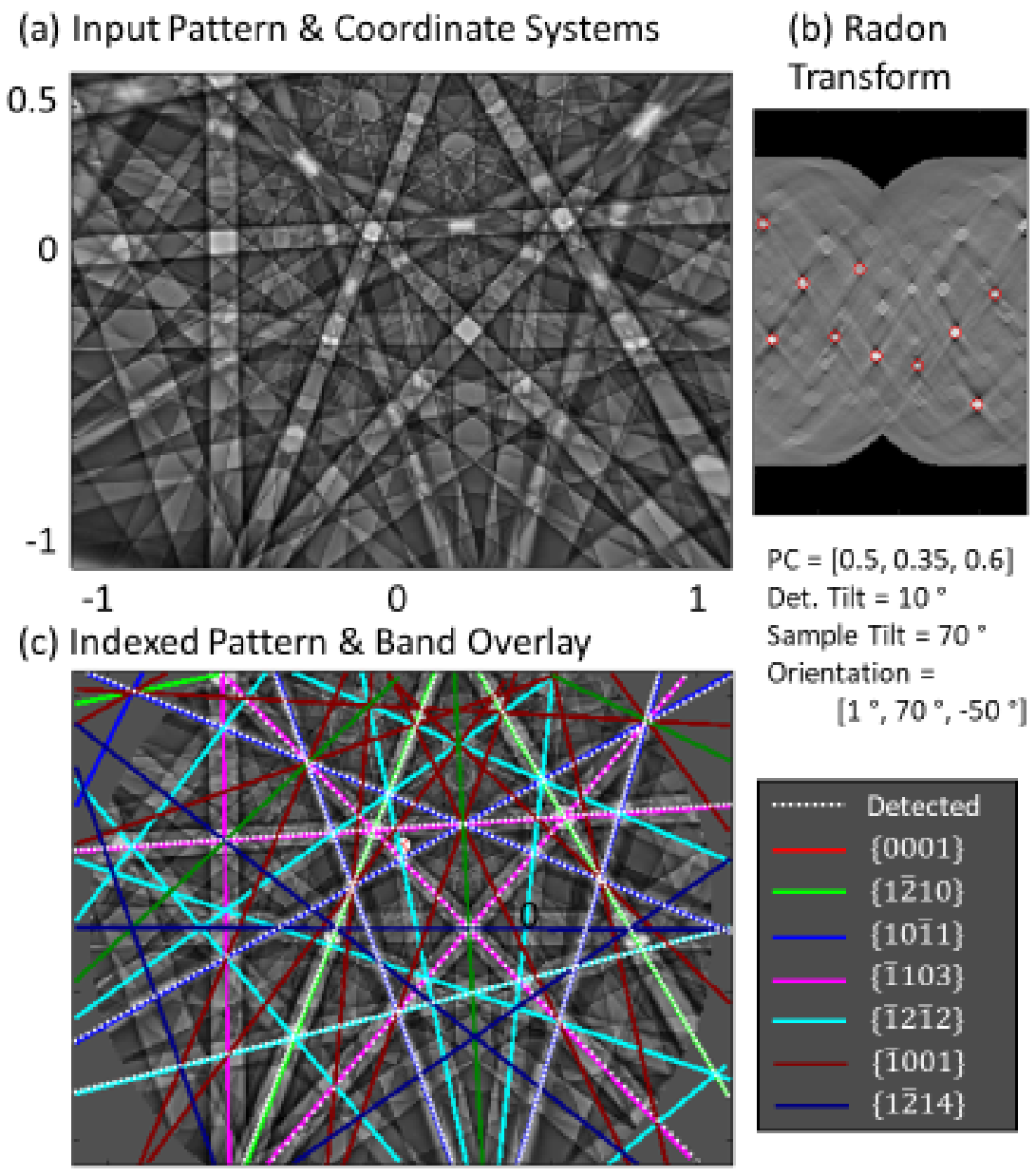

**Figure 5** Dynamically simulated HCP titanium diffraction pattern, with exact pattern centre location known *a priori*. Ten bands, identified using the radon transform, were used in the indexing routine.

We test whether the pattern centre can be found with the WMAE approach using a pattern with a known pattern centre and indexing with an incorrect pattern centre of [0.6, 0.35, 0.6]. Figure 6 shows the MAE, number of correctly indexed bands, and the WMAE as a function of assumed pattern centre position. The MAE fields (Figure 6) shows a minimum of the WMAE which is located at the exact pattern centre, and this minimum is smooth within ±0.025 of the correct pattern centre. Use of the WMAE ensures that the solution is strongly localised near the accurate solution (as in the limit, 3 'indexed' bands will have an exact solution and so zero MAE).

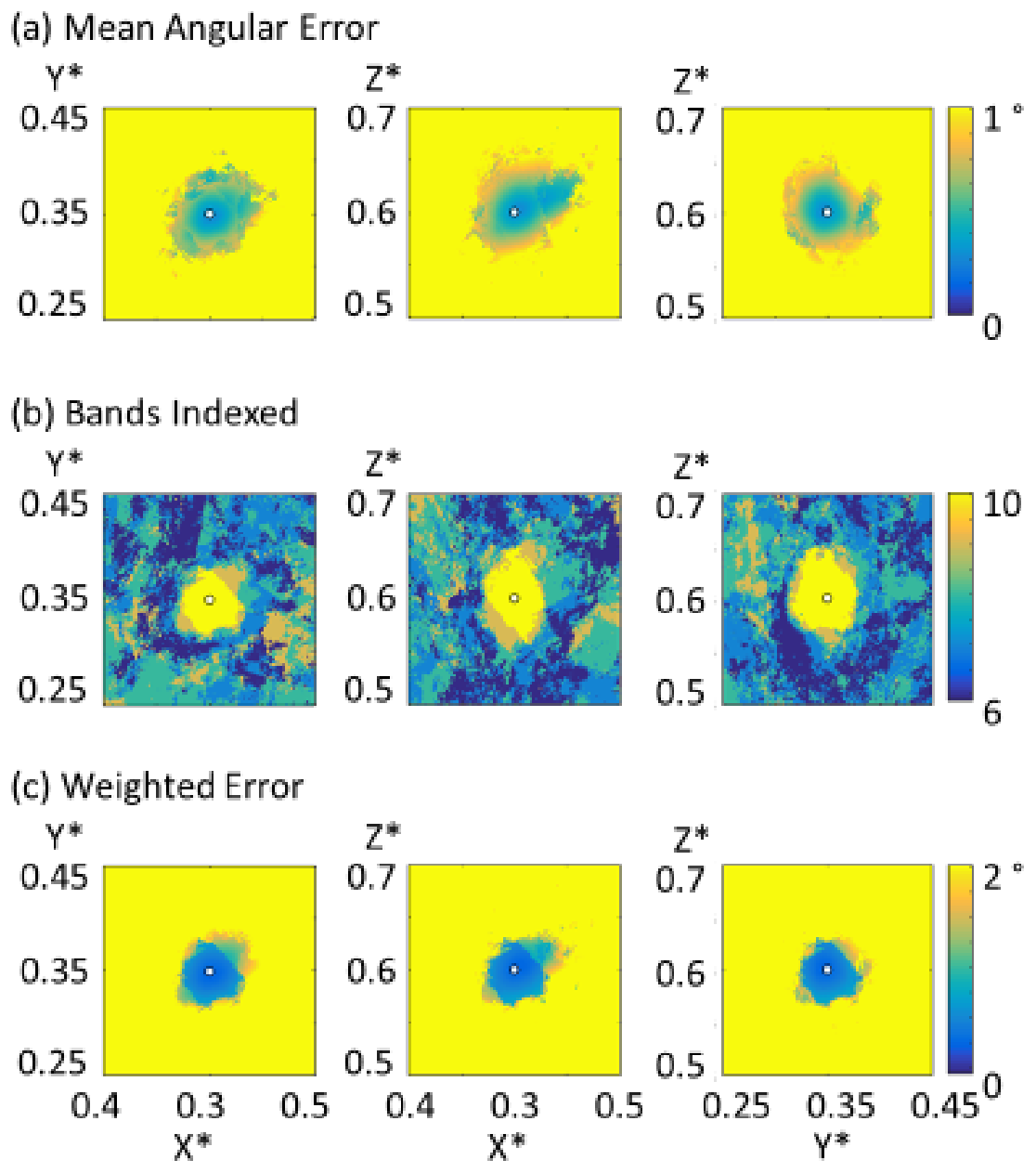

**Figure 6** Pattern indexing of a simulated pattern, which has a known pattern centre of [0.5, 0.35, 0.6] . Indexing is performed with displaced pattern centre, and the error metrics are calculated: (a) mean angular error; (b) number of bands, and (c) weighted error. In practice we search for a solution which has a minimum in the weighted mean angular error plot.

## 4. Accuracy Determination

We test the accuracy of this indexing methodology using dynamical simulations. A 'master' EBSP was generated in a stereographic projection, and from this master EBSP patterns with different detector positions and crystal orientations were generated using a bi-cubic interpolation of the intensity variation (an example is shown in Figure 6). We imported these simulated patterns and indexed them with this new indexing methodology.

Accuracy of the pattern centre determination method was performed using 200 generated patterns at 400x300 pixel resolution, with different crystal orientations, and pattern centres as 400x300 pixel patterns. The pattern centre was varied [0.5 ± 0.1, 0.35 ± 0.1, 0.6 ± 0.1] with uniform spacing. The recorded pattern centre was determined using the optimisation procedure, and accuracy was determined. Accuracy of this pattern centre determination was evaluated with the following settings: $\theta = 1°$, $\rho = 150$, and 10 peaks; and identification with a LUT containing a diffracting lattice plane (i.e. reflector) list with {0 0 1}; {1 -2 0}; {1 0 1}; {-1 1 3}; {-1 2 2}; {0 -2 1}; {-3 1 1}; {1 0 0}; {1 -2 4}.

The results are shown in Figure 7 and the absolute error of the determined pattern centre was found to be within 0.01 (in pattern fractions).

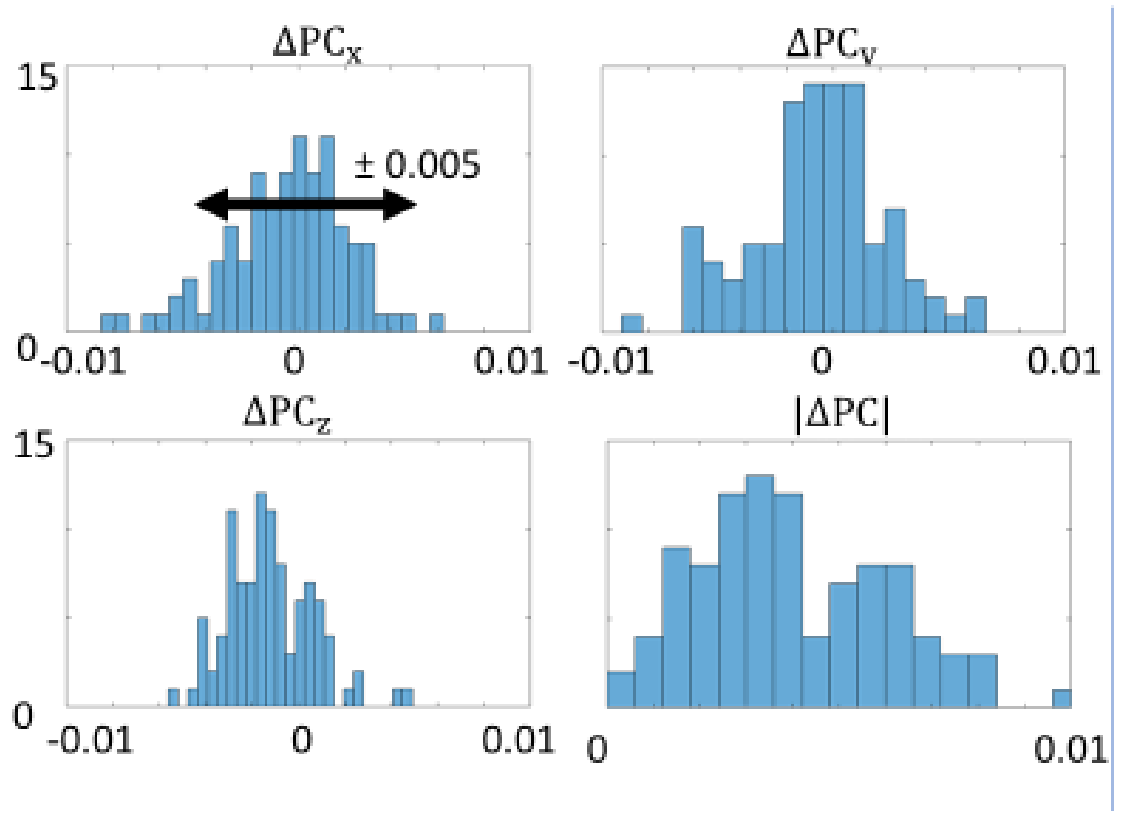

**Figure 7** Accuracy assessment of the pattern centre determination, using random orientations and random pattern centres and dynamical simulations of diffraction patterns.

We determine the accuracy of orientation measurement using patterns with a known (and fixed) pattern centre of [0.5, 0.3, 0.6] and 1000 patterns with orientations sampled from a near uniform sampling of the SO(3) space for the titanium symmetry operators. The accuracy of indexing with these settings is better than 0.5° for the settings listed above, as shown in Figure 8.

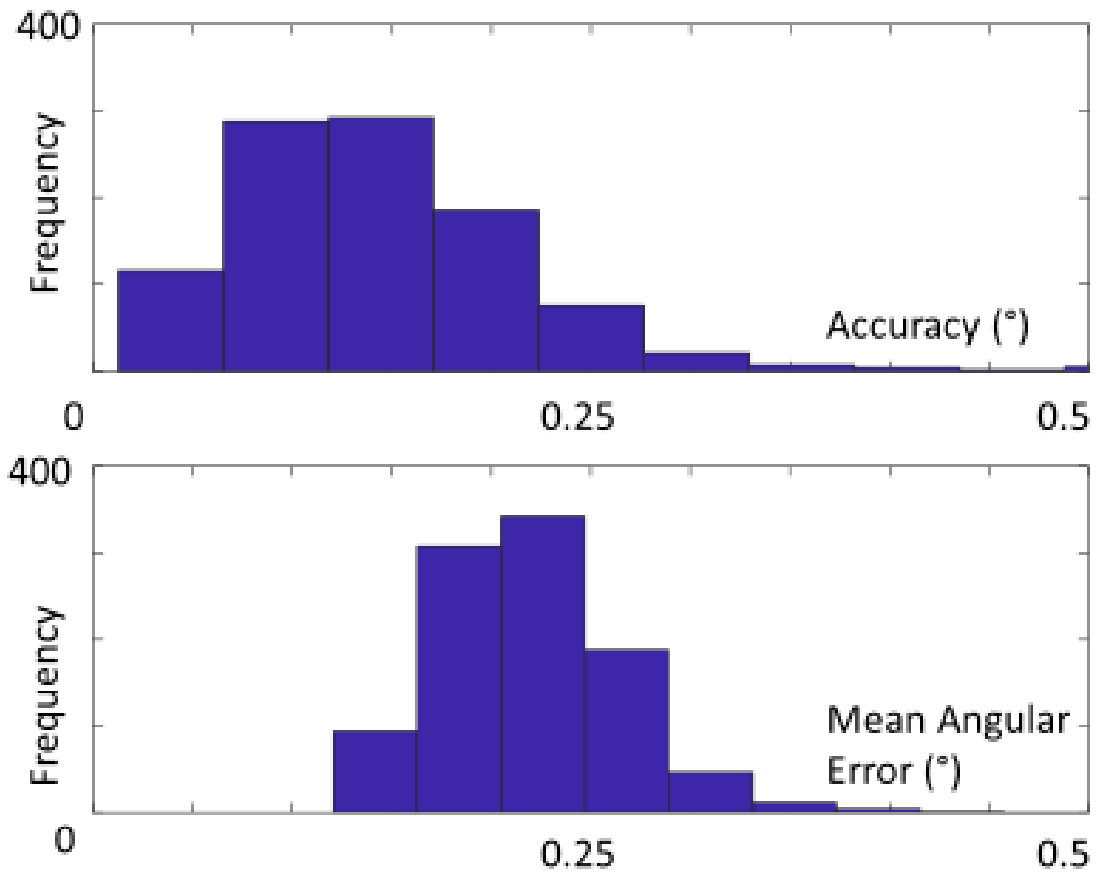

**Figure 8** Accuracy assessment for indexing dynamical patterns.

### 5. Demonstration – Experimental Patterns

An experimental EBSP from a single crystal (001) Si wafer was captured using a 20 keV electron probe in a FEI Quanta instrument equipped with a Bruker eFlash HR EBSD detector. The raw pattern was captured at full resolution of the detector of 1200 x 1600 pixels (cropped in Y slightly to 1152). A greyscale bit depth of 12 (saved in 16 bit containers) was used. We have successfully determined the pattern centre and indexed this pattern, as shown in Figure 9.

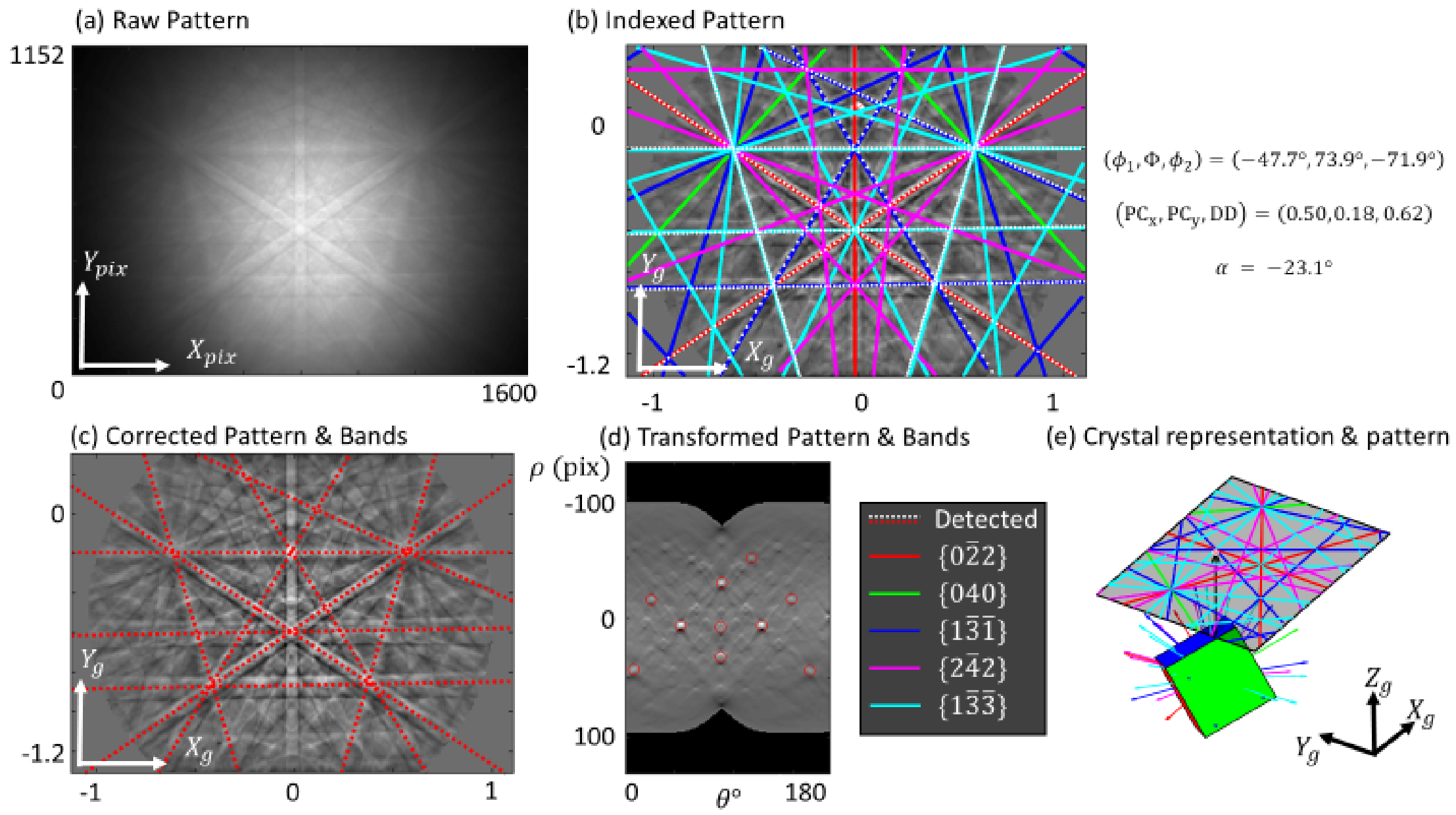

**Figure 9** Demonstration of indexing for a silicon diffraction pattern: (a) Raw uncorrected pattern, in pixel coordinates; (b) pattern in gnomonic coordinate system, with bands identified & indexed and pattern centre shown; (c) Pattern as corrected using a low frequency filter and circular crop, with identified bands overlaid; (d) Radon transform of background correct pattern, with band centres identified; (e) 3D representation of crystal and pattern.

To show that this technique is appropriate for maps, we have also used the technique to index a sample of interstitial free (IF) steel which has been plastically deformed. The results of this indexing are shown in Figure 10. The pattern centre was determined for multiple points in a (sub-sampled) grid in this 2D map and a planar pattern centre model was applied to index each point with an appropriate pattern centre.

To assess the effectiveness of this indexing approach, we introduce several quality metrics inspired by prior EBSD methods:

1. Pattern quality describes the intensity of the identified peaks with respect the flat fielded Radon transformation. This is useful to highlight grain boundaries, surface roughness and phase.
2. Slope highlights grain boundaries, and is calculated through the sum of the quotient of the positive and negative edges of each Radon peak.
3. The number of bands used in indexing, based upon a threshold of 3° in the refinement step.
4. The mean angular error (see Equation 6).

The IPF-maps in Figure 10 show good agreement between analysis in the commercial Bruker ESPRIT software (v2.1), with up to 12 bands indexing and automatic Hough settings of '60', and our open source AstroEBSD MATLAB code.

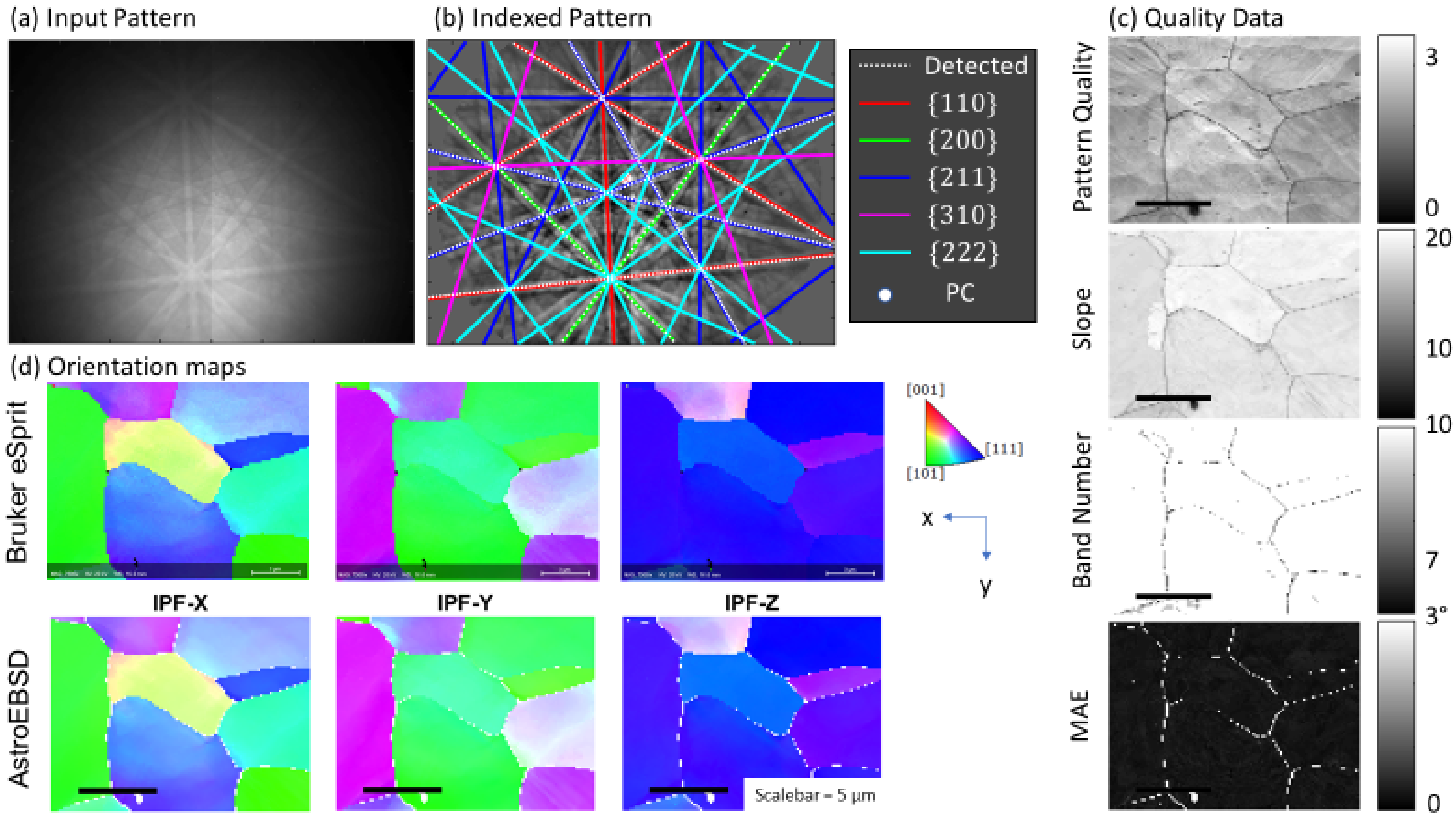

**Figure 10** Demonstration on an IF steel (α-iron) experimental map. (a) example input pattern; (b) indexing of this pattern, with unknown pattern centre; (c) quality maps; (d) comparison of Bruker ESPRIT orientation maps and AstroEBSD maps. Performance of AstroEBSD is comparable to the Bruker ESPRIT indexing, with the exception of the grain boundary pixels.

### 6. Discussion

We present a new algorithm for indexing EBSD patterns. This algorithm relies significantly upon prior work, principally knowledge of suitable image processing strategies used to index diffraction patterns. A reasonably robust methodology for pattern indexing has been presented and released at Open Source software and this new pattern indexing approach assesses the interplanar angle information encoded in a series of characteristic triangles. The potential of this indexing approach is highlighted principally through the first accuracy assessment of Hough/Radon based EBSP indexing, where dynamical diffraction simulations of controlled orientations and pattern centres are indexed. The results are systematically compared with the input pattern centres and orientations to establish an accuracy limit for these pattern analysis settings.

The performance of our open source algorithm is demonstrated in this test case on an IF steel experimental map (Figure 10) and comparison with the commercial indexing indicates that the AstroEBSD performance is comparable. Ultimately we do not know which one is best in these cases, as the "true" orientation is uncertain. However we do note that we have filtered orientations near the grain boundary in the AstroEBSD case, due to their high MAE, more strongly than our reported values from the Bruker software. We anticipate that this is most likely due to pattern overlap issues which makes band localisation and extraction more difficult, especially for our simple Sobel edge based band localisation approach. In cases where EBSD analysis is to be performed near or on grain boundaries, this would be an area where the community could easily adjust this step within the AstroEBSD indexing routine and upload modifications as "branches" within the github repository.

The algorithm builds upon the initial star cataloguing work of Groth (Groth, 1986) and it is likely that there are significant algorithm improvements that can assist in speeding up this process. Indexing can occur at ~50 Hz with the present algorithm (on a single core of a Xenon CPU E5-2600 v3 @ 2.60GHz) and it is likely that inter-disciplinary approaches, such as borrowing more from the astronomy community (Spratling & Mortari, 2009), may assist here.

We do not claim that the accuracy reported here is an ultimate accuracy (and it may not translate well to commercial packages), and we anticipate that there will be methods to significantly improve the accuracy of Radon based indexing. Ultimately, the accuracy of the indexing method relies significantly on determination of the plane normal vectors, which requires accurate localisation of the band centres (if a Radon/Hough space routine is to be employed). In this work, we have introduced a simple method of finding the band centres which used an edge detection algorithm in the Radon transform, searching along the $\rho$ direction. If the diffraction bands are sharp, then this provides a reasonable and simple edge detection algorithm. However, this step could be optimised (e.g. as found in one commercial vendor and utilising ideas within the 3D Hough (Maurice & Fortunier, 2008)) and further optimisation could be based upon a successful first pass 'index' and then utilisation of more bands, or improved localisation, to refine the orientation (remaining within a Radon based schema). In particular, optimisation of the peak centre localisation would likely increase the accuracy of the Radon based method. Further optimisations could include tackling pseudo-symmetry issues (e.g. in geological samples) and we hope that in releasing this software as open source we can promote development of transparent indexing algorithms. We anticipate that this could be an exciting and useful field to explore.

We have started to perform accuracy assessments of this approach, demonstrated here with pattern centre determination and crystal orientation determination using simulated patterns. We present these results with care:

1. They are obtained from high quality and sharp dynamical diffraction patterns.
2. There are limited pseudo-symmetry issues in these simulations.
3. Accuracy artefacts due to scan precision, detector positioning, and sample cutting, will make the determination of absolute orientation more challenging (Nolze, 2007).

4. The precision of the pattern centre determination assumes that the crystal is unstrained. It is well known that this is problematic for use in attempts towards absolute strain measurement with high angular resolution approaches (Alkorta, 2013, Britton *et al.*, 2010).

Recently, we have been following the new template matching based ‘pattern indexing’ approaches, such as developed by de Graef and co-workers (Chen *et al.*, 2015, Ram *et al.*, 2017, Singh & De Graef, 2017), where by a series of good quality simulations are template matched against experimental patterns and the best match is selected as the indexed solution. This approach may prove useful, and complimentary, to other approaches but the generation of a significant number of high quality templates is computationally expensive and so it is likely that Hough or Radon based image processing schemes will continue to be of value.

## 7. Conclusion

We have presented a new indexing algorithm, which is released as open source software. This algorithm assists in determining ‘which band is which’ in EBSD pattern analysis, based upon *a-priori* knowledge of the crystal phases and we have included image processing steps to assist in this process. This new algorithm is demonstrated with a few case studies, refinement of the pattern centre to reasonable precision, within 0.5% of accuracy in the case of dynamical patterns, and determination of the orientation with an accuracy of >0.25° can be achieved (based upon indexing of the pattern, and not due to uncertainties in sample and detector alignment). Finally, we demonstrate that this method can be used on a deformed IF steel polycrystal and that the orientation maps achieved are commensurate with those resolved by a commercial indexing package. We hope that this algorithm will boldly open up new enterprises in EBSD pattern indexing.

## 8. Acknowledgements

We acknowledge helpful initial discussions with Austin Day. We thank Aimo Winkelmann for helpful discussions on indexing approaches.

## 9. Author Contributions

TBB developed the initial indexing routine while in Oxford working with AJW. Many bugs were addressed when VT adapted the code for experimental patterns. The dynamical simulations and analysis was performed by AF. The iron example map was captured by JH and TBB. The initial manuscript was drafted by TBB, with all authors contributing to the final version. The AstroEBSD software tool was completed at Imperial College London.

## 10. Data Statement

The iron experimental data for this paper can be downloaded from a Zenodo repository. An up-to-date copy of AstroEBSD can be downloaded from GitHub (https://github.com/benjaminbritton/AstroEBSD/).